\documentclass[manuscript]{aastex}
\usepackage{graphicx}

\newcommand{\ms}{\noalign{\vspace{3pt plus2pt minus1pt}}}

\newcommand{\be}{\begin{equation}}
\newcommand{\ee}{\end{equation}}
\newcommand{\bea}{\begin{eqnarray}}
\newcommand{\eea}{\end{eqnarray}}

\newcommand{\gapprox}{\lower.4ex\hbox{$\;\buildrel >\over{\scriptstyle\sim}\;$}}
\newcommand{\lapprox}{\lower.4ex\hbox{$\;\buildrel <\over{\scriptstyle\sim}\;$}}

\def\rmd{d}
\def\rme{e}
\def\rmi{i}
\def\bi{\bf}

\shorttitle{Linear acceleration emission 1}
\shortauthors{Melrose, Rafat and Luo}

\begin{document}

\title{Linear acceleration emission: 1~Motion in a large amplitude electrostatic wave}

\author{D. B. Melrose, M. Z. Rafat and Q. Luo}

\affil{Sydney Institute of Astronomy, School of Physics, University of Sydney, NSW 2006, Australia}

\begin{abstract}
We consider the motion of a charge in a large amplitude electrostatic wave with a triangular wave form relevant to an oscillating model of a pulsar magnetosphere. The (one-dimensional) orbit of a particle in such a wave is found exactly in terms of Weierstrass functions. 

The result is used to discuss linear acceleration emission (at both low and high frequencies) in an oscillating model for pulsars. An explicit expression for the emissivity is derived in an accompanying paper \citep{ml09}, and used here to derive an expression for the absorption coefficient at low frequencies. We show that absorption can be negative, corresponding to maser emission. For the large amplitude required to trigger pair creation in an oscillating model, the rate of the maser growth is too small to be effective. Effective growth requires smaller amplitudes, such that the maximum Lorentz factor gained by acceleration in the wave is $\lapprox10$.
\end{abstract}


\section{Introduction}

An oscillatory model for a pulsar magnetosphere introduces an emission mechanism that depends explicitly on the oscillating electric field: linear acceleration emission (LAE). Our purpose in this paper, and in an accompanying paper \citep{ml09}, referred to hereinafter as paper~2, is to develop the theory for LAE in a large amplitude electrostatic wave (LAEW), and discuss its viability as a pulsar emission mechanism. 

In polar-cap models for pulsars, the inductive electric field associated with the rotation of the magnetized neutron star cannot be completely screened everywhere, and the residual field in `gaps' leads to acceleration of particles along the field lines to high enough energies to trigger a pair cascade. The polar-cap region is populated by the resulting secondary pairs, which escape through the light cylinder to form the pulsar wind. In most models, the electric field, the acceleration, the pair creation, and the screening of the inductive field are assumed to occur in a time-stationary way in a frame corotating with the star. An alternative view is that these processes occur in an oscillatory way \citep{s71,ake75}. Recent numerical models confirm that the system is unstable to temporal perturbations that develop into large amplitude electrostatic oscillations, both for ordinary pulsars \citep{letal05,metal05} and for magnetars \citep{bt07a,bt07b}. In these numerical models the oscillations are purely electrostatic and purely temporal. \cite{b08} proposed Alfv\'en waves as an alternative to the electrostatic oscillations, and \cite{lm08} proposed an analytic model that generalized the purely temporal oscillations to large amplitude propagating waves. The analytic model includes both purely temporal and purely spatial oscillations as limiting cases; purely spatial oscillations were found by \cite{s97}. More generally, the analytic model describes an outward-propagating, large-amplitude electrostatic wave; the phase velocity, $c\beta_V$, is a free parameter with purely temporal oscillations corresponding to $\beta_V\to\infty$, and purely spatial oscillations to $\beta_V=0$. A bulk outward motion occurs at the counterpart of the group velocity, which corresponds to a speed $c/\beta_V$ in the superluminal  case, $\beta_V>1$. For $\beta_V>1$ one can make a Lorentz transformation to the frame moving at $\beta^*=-1/\beta_V$, referred to here as the primed frame, in which the oscillations are purely temporal, and it is convenient to adopt this frame for detailed calculations. We envisage a model for a pulsar magnetosphere in which the pair creation occurs in sporadic, isolated events association with the development of LAEWs. It is possible that such a model, when averaged over the effects of the LAEWs, mimics a conventional time-stationary model, but it is more likely that new features introduced by the LAEWs result in important differences from conventional time-stationary models, with the emission of LAE being one such new feature.

Pair creation is included in the oscillating model, but has not been treated in the same detail as in the stationary model \citep{zh00,ha01,ae02}.  The rate of pair creation increases with the amplitude of the LAEW, providing a mass loading that limits the amplitude \citep{letal05}. We assume that the pair creation produces particles with a spread in Lorentz factors that is small compared with the maximum Lorentz factor, $\gamma_{\rm max}$, achieved in the LAEW, and we neglect this spread for the `background' particles that are described in terms of their (oscillating) velocity, Lorentz factor and number density. The results derived here are not sensitive to the frequency of the LAEW, which we estimate to be $\sim10^8$--$10^9\rm\,s^{-1}$, based on $\omega_p/\gamma_{\rm max}^{1/2}$ \citep{letal05}, with $\gamma_{\rm max}=10^6$--$10^7$ and the plasma frequency determined by the Goldreich-Julian number density times a multiplicity $M$: $\omega_p\approx\sqrt{2\Omega_e\Omega M}\sim 5\times10^{11}\rm\,s^{-1}$ for $M=100$, a cyclotron frequency, $\Omega_e$ for $B=10^9\,$T, and a rotation frequency, $\Omega=2\pi/P$ with $P=0.1\,$s.

In the present paper, we discuss the motion of a charge in a LAEW, and the associated emission of radiation. A first integral of the equation of motion is obtained for a particle in a LAEW with an arbitrary wave form. A second integral requires a specific choice of wave form. The solution for a sinusoidal wave form was found by \cite{r92a,r92b} in terms of elliptic functions. Here we consider a sawtooth or triangular wave form, which is an excellent approximation to the actual wave form for a LAEW \citep{lm08} in the case where the particles are highly relativistic \citep{lm08}. The solution for the orbit  is found in terms of Weierstrass elliptic integrals. The exact solution is used as the basis for a treatment of LAE in a LAEW, which is developed in terms of Airy integrals in paper~2, and used here to discuss the possibility of maser LAE in a LAEW as a possible pulsar radio emission mechanism. 

The equation of motion for a relativistic particle is solved in two superficially different ways. In Appendix~\ref{sect:covariant}, a covariant form for the wave equation is used to derive a solution in terms of (Lorentz) invariants; the solution may be written in any inertial frame by expressing the invariants in terms of quantities in that frame. In Sec.~\ref{sect:pframe} the solution is obtained in the primed frame, defined to be the frame in which the oscillations in the LAEW are purely temporal. (There is only a notational difference between the important results in Appendix~\ref{sect:covariant} and Sec.~\ref{sect:pframe}.) The exact orbit for a triangular wave form is found in Sec.~\ref{sect:triangular}. The application to LAE is discussed in Sec.~\ref{sect:LAE}, concentrating on the possibility of maser LAE at low frequencies. The results are discussed in Sec.~\ref{sect:discussion}, and our conclusions are summarized in Sec.~\ref{sect:conclusions}.

\section{Motion in a LAEW: primed frame}
\label{sect:pframe}

In this section we find the orbit of a charge in a LAEW in the primed frame. First, we describe the transformation between the inertial frames, and write down the forms of the invariants introduced in Appendix~\ref{sect:covariant} in the primed frame. The motion is assumed to be restricted to one dimension (the 3-axis), along the direction of the superstrong magnetic field in a pulsar magnetosphere.

\subsection{Description of the LAEW}

The LAEW is described in terms of its amplitude, $E(\chi)=E_0T(\chi)$, is a function of phase, $\chi$, with the maximum electric field, $E_0$, written in terms of a frequency, $\omega_E$, and with the wave form described by the function $T(\chi)$, with maxima and minima $\pm1$:\\
$\eta\to\epsilon$ everywhere
\be
E(\chi)=\epsilon{m\omega_E\over q}T(\chi),
\qquad
\omega_E={|q|E_0\over mc},
\label{Tchi}
\ee
where the sign, $\epsilon=q/|q|$, is opposite for electrons and positrons. For an arbitrary wave form, the only specific assumption made is that $T(\chi)=T(\chi+2\pi)$ is periodic. A LAEW, even with a non-sinusoidal wave profile, is described in terms of its frequency, $\Omega$, and its wave vector, $K$, along the 3-axis. The LAEW is assumed superluminal, implying that its phase velocity $\Omega/K=\beta_Vc$, satisfies $\beta_V>1$. 

\subsection{Lorentz transformation}

The Lorentz transformation to the primed frame corresponds to a boost with velocity (in units of $c$) $\beta^*=-1/\beta_V$ and Lorentz factor $\gamma^*=\beta_V(\beta_V^2-1)^{-1/2}$. The relation between time and position in the two frames is
\be
ct'=\gamma^*(ct-\beta^* z),
\qquad
z'=\gamma^*(z-\beta^*ct).
\label{LT1}
\ee
In the primed frame, the frequency, $\Omega'$, of the oscillation and the 4-velocity, $u^{\mu'}=\gamma'[1,0,0,\beta']$ of the test charge are given by
\be
\Omega'={\Omega\over\gamma^*},
\qquad
\gamma'=\gamma^*\gamma\left(1-{\beta\over \beta_V}\right),
\label{LAE2}
\ee
where the wave vector in the primed frame is zero by construction, $K'=0$. The frequency and angle of emission of LAE are related in the two frames by
\be
\omega=\gamma^*\omega'(1+\beta^*\cos\theta'),
\qquad
\cos\theta={\cos\theta'+\beta^*\over1+\beta^*\cos\theta'}.
\label{LT2}
\ee

Relevant invariants introduced in Appendix~\ref{sect:covariant} have the following values in the primed frame, implied by $K^{\mu'}=(\Omega',0,0,0)$, $K_D^{\mu'}=(0,0,0,\Omega')$:
\be
\Omega'=(K^2)^{1/2},
\quad
\chi=\Omega't',
\quad
K{\tilde u}(\chi)=\Omega'\gamma',
\quad
K_D{\tilde u}(\chi)=-\Omega'u',
\label{invariants}
\ee
where we assume $K^2=\Omega^2-|{\bi K}|^2>0$ and write $u'=\gamma'\beta'$. The electric field, $E(\chi)$, is an invariant, and has the same form in all frames.

\subsection{Equation of motion}

The equation of motion may be written in two equivalent forms:
\be
{\rmd\gamma'\over\rmd\chi}={\omega_E\over\Omega'}T(\chi)\,u',
\qquad
{\rmd u'\over\rmd\chi}={\omega_E\over\Omega'}T(\chi)\gamma'.
\label{LAE3}
\ee
Integrating the second of (\ref{LAE3}) gives
\be
u'(\chi)=u'_0+(\omega_E/\Omega')F(\chi),
\label{LAE4}
\ee
where $u'_0=\gamma'_0\beta'_0$ contains the initial conditions, and with $F(\chi)$ given by (\ref{LAE10a}), with $a_0=0$ assumed in (\ref{LAE6}). The solution (\ref{LAE4}) implies a Lorentz factor
\be
\gamma'(\chi)=\{1+[u'_0+(\omega_E/\Omega')F(\chi)]^2\}^{1/2},
\label{LAE5a}
\ee
and a velocity (in units of $c$)
\be
\beta'(\chi)={u'_0+(\omega_E/\Omega')F(\chi)\over
\{1+[u'_0+(\omega_E/\Omega')F(\chi)]^2\}^{1/2}}.
\label{LAE5}
\ee

\subsection{Background and test charges}

In the solution (\ref{LAE4}), the particle has an arbitrary initial velocity, $\beta'_0c$, at $\chi=0$. It is convenient to separate the particles into two classes: background particles and test charges. The background particles are an idealized class of electrons and positrons that move in opposite directions in phase with the LAEW, with $\beta'_0$ assumed to be identically zero. In the idealized case where there is no intrinsic velocity spread in the background particles, such particles are instantaneously at rest twice per period, and reach their maximum Lorentz factor twice per period. Let the initial phase, $\chi=0$, be chosen such that the background particles have $\beta'_0=0$ in (\ref{LAE4}), so that they are instantaneously at rest at $\chi=n\pi$ for any integer $n$. The background particles have their maximum Lorentz factors, $\gamma'_{\rm max}$, at the phases $\chi=\pi/2+n\pi$. A test charge is then any particle that has $\beta'_0\ne0$ at $\chi=0$. In the analytic model \citep{lm08} all particles are in effect test charges. The concept of a background particle is useful when considering situations in which the velocity spreads of the electrons or of the positrons are unimportant, and are approximated by zero.

The maximum 4-velocity for the background particles, $u'_{\rm max}=\gamma'_{\rm max}\beta'_{\rm max}$, is determined by (\ref{LAE4}), which becomes
\be
u'(\chi)=u'_{\rm max}{F(\chi)\over F(\pi/2)},
\qquad
u'_{\rm max}={\omega_E\over\Omega'}F(\pi/2).
\label{LAE4a}
\ee
The Lorentz factor corresponding to (\ref{LAE4a}) is
\be
\gamma'(\chi)=\big\{1+
(\omega_E/\Omega')^2[F(\chi)]^2\big\}^{1/2}.
\label{LAE4b}
\ee

\begin{figure*}
\centerline{
\includegraphics[width=10cm]{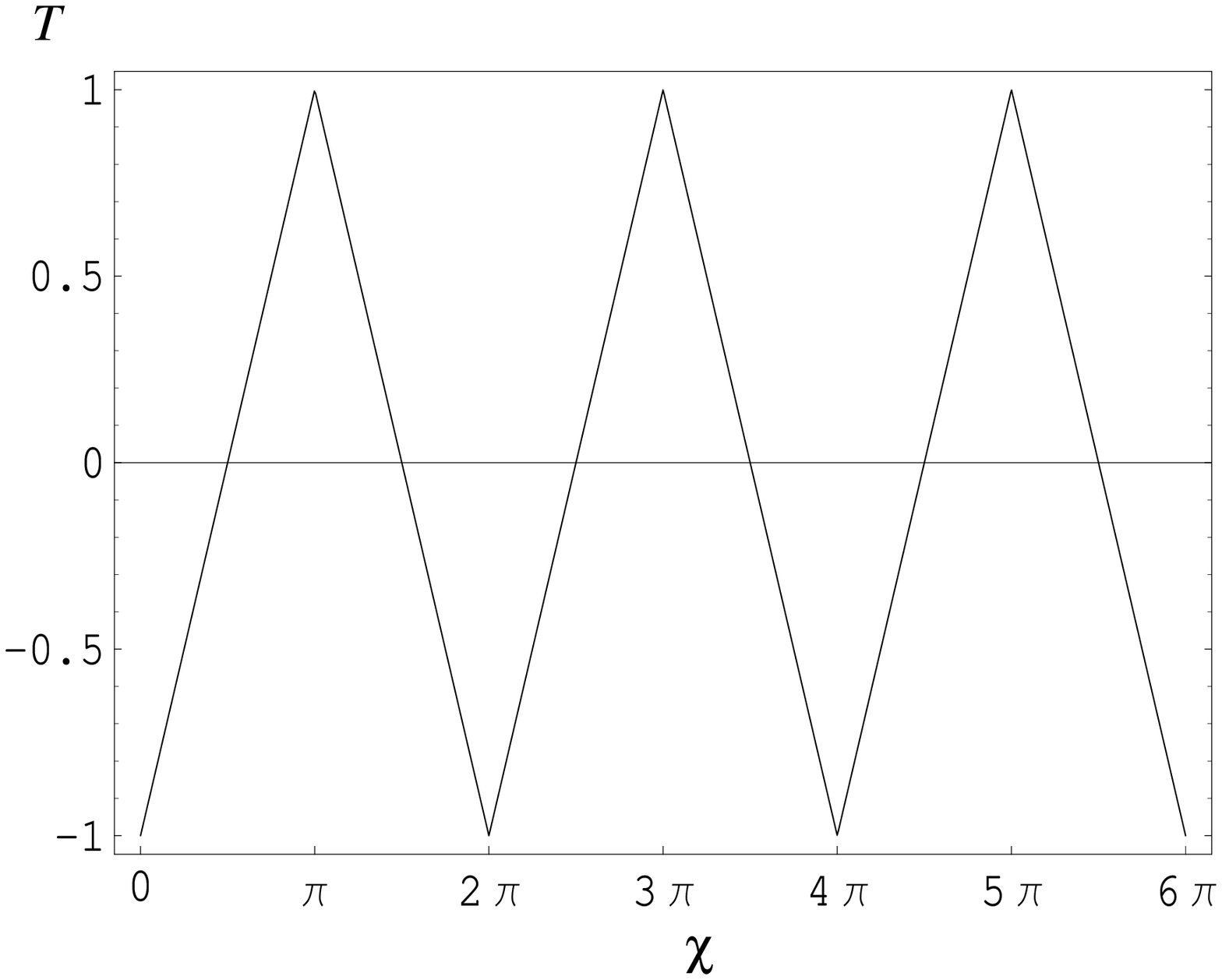}
}
\caption{The triangular wave form is illustrated for a LAEW; in the exact theory \citep{lm08}, the sharply curved parts correspond to a range of phase $\sim1/\gamma'_{\rm max}$ where the background particles are briefly nonrelativistic.}
\label{fig:efield}
\end{figure*}

\subsection{Orbit}

A second integration gives the orbit:
\be
z'=z'_0+{c\over\Omega'}Z'(\chi),
\qquad
Z'(\chi)=\int_0^\chi\rmd\chi'\,\beta'(\chi').
\label{TW3}
\ee
The explicit form for $Z'(\chi)$ obtained by inserting the solution (\ref{LAE5}) into (\ref{TW3}) cannot be evaluated in closed form in general. It is evaluated for a triangular wave form in Sec.~\ref{sect:triangular}.

The motion implied by (\ref{TW3}) for a background particle, $u'_0=0$, is an oscillation around a fixed point; the particle returns to the same position at the same phase in each subsequent oscillation, and its mean velocity (and 4-velocity), averaged over a period of the LAEW, is zero. A test charge, $u'_0\ne0$, has a net drift motion, in the sense that its velocity (and its 4-velocity) averaged over a period is nonzero. One may regard a test charge as oscillating about a drifting point.  For $\gamma'_0<\gamma'_{\rm max}$, a test charge is instantaneously at rest twice per period; for $\gamma'_0>\gamma'_{\rm max}$, $\beta'(\chi)$ never changes sign.

\section{Triangular wave form}
\label{sect:triangular}

For $\omega_E/\Omega'\gg1$ the electric field is strong enough to cause all particles to be highly relativistic for all phases of the LAEW with the exception of a brief phase where the direction of their velocity reverses. The LAEW is then well approximated by the triangular wave form \citep{lm08}, illustrated in Fig.~\ref{fig:efield}. An interpretation of the  triangular wave form follows by noting that the current density, ${\bi J}$, associated with relativistically counterstreaming electrons and positrons is approximately unchanged by the acceleration (by ${\bi E}$ parallel to ${\bi J}$), and that the solution of $\partial {\bi E}/\partial t=-{\bi J}/\varepsilon_0$ implies that ${\bi E}$ varies linearly with $t$; ${\bi J}$ reverses sign abruptly in the brief phases where the particles are nonrelativistic, and ${\bi E}$ reverses sign out of phase with ${\bi J}$. Here we solve explicitly for the orbit for such a triangular wave form.

\subsection{Form for $T(\chi)$ and $F(\chi)$}

Our model for a triangular wave form is
\be
T(\chi)=\left\{
\begin{array}{ll}
2\chi/\pi-1,
& 0<\chi<\pi,
\\
\ms
3- 2\chi/\pi,
& \pi<\chi<2\pi.
\end{array}
\right.
\label{TW1}
\ee
The function $F(\chi)$, defined by (\ref{LAE10a}), is
\be
F(\chi)={\epsilon\over\pi}
\left\{
\begin{array}{ll}
(\chi-\pi/2)^2-\pi^2/4,
& 0<\chi<\pi,
\\
\ms
-(\chi-3\pi/4)^2+\pi^2/4,
& \pi<\chi<2\pi.
\end{array}
\right.
\label{TW2}
\ee
A background particle has extrema in its velocity at $\chi=\pi/2+n\pi$, so that (\ref{LAE4a}) implies
\be
|u'_{\rm max}|=\pi\omega_E/4\Omega',
\label{LAE4c}
\ee
with $|u'_{\rm max}|\approx\gamma'_{\rm max}\gg1$ in the case of most interest here.

\subsection{Exact solution for the trajectory}
\label{sect:weierstrass}

The integral over $\chi'$ in (\ref{TW3}) with (\ref{LAE5}) may be carried out exactly for the triangular wave form, in terms of elliptic integrals, specifically, the Weierstrass $\wp(x)$ and $\zeta(x)$ functions \citep{AS65}. The integral is written as a sum over integrals of the form
\be
I=\int_{\chi_1}^{\chi_2}\rmd\chi\,{(\chi-a)^2-p\over
\sqrt{[(\chi-a)^2-p]^2+b}},
\label{int}
\ee
where $\chi_1,\chi_2,a,p$ and $b$ are constants. By making the substitution $y=(\chi-a)^2-2p/3$, the integral (\ref{int}) reduces to the form
\be
I=\pm\int_{y_1}^{y_2}\rmd y'\,{y'-p/3\over
\sqrt{4y'^3-g_2y'-g_3}},
\label{int2}
\ee
with $g_2=4(p^2-3b)/3$, $g_3=-8p(p^2+9b)/27$, which is a standard form for the Weierstrass functions \citep{AS65}
\be
\zeta(x;g_2,g_3)=-\int dx\,\wp(x;g_2,g_3),
\label{zeta}
\ee
denoted $\zeta(x)$ here, with $\wp(x;g_2,g_3)$, denoted $\wp(x)$ here, defined implicitly by $y=\wp(x)$ and
\be
x=\int_\infty^ydy'{1\over\sqrt{4y'^3-g_2y'-g_3}}.
\label{wp}
\ee

Two sets of choices, $a_i,p_i$ with $i=1,2$ are needed to find the full solution, and these define
\be
x_i(\chi)=\wp^{-1}\big((\chi-a_i)^2-2p_i/3\big).
\label{xi}
\ee
In terms of the functions
\be
w_i(\chi)=p_ix_i(\chi)+3\zeta\big(x_i(\chi)\big),
\label{wi}
\ee
with $g_{2i},g_{3i}$ evaluated for $p=p_i$, the solution is
\be
Z'(\chi)={\epsilon\over3}
\left\{
\begin{array}{cc}
-w_1(0)+w_1(\chi),&0<\chi<\pi/2,\\
w_1(\pi/2)-w_1(\chi).&\pi/2<\chi<\pi,\\
w_2(\pi)-w_2(\chi),&\pi<\chi<3\pi/2,\\
-w_2(3\pi/2)+w_2(\chi),&3\pi/2<\chi<2\pi,
\end{array}
\right.
\label{Zchi}
\ee
with $a_1=\pi/2$, $p_1=\pi^2/4-\pi\gamma'_0\beta'_0/u'_{\rm max}$, $a_2=3\pi/2$, $p_2=\pi^2/4+\pi\gamma'_0\beta'_0/u'_{\rm max}$, $b=\pi^2/u'^2_{\rm max}$.

\begin{figure*}
\centerline{
\includegraphics[width=0.5\textwidth]{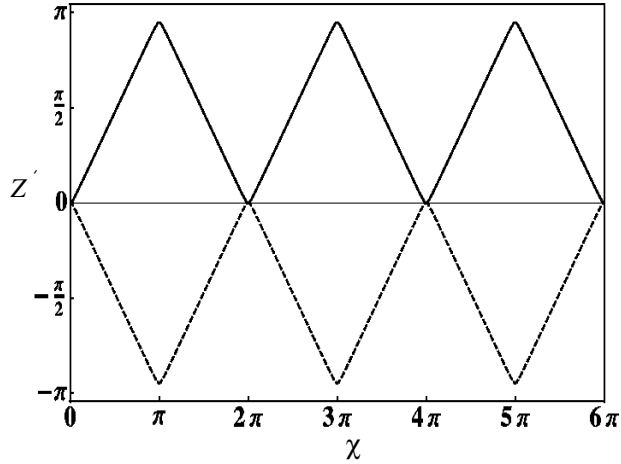}
}
\caption{The trajectory of a background particle (solid curve for an electron, dashed curve for a positron) in a LAEW for a modest relativistic value, $u'_{\rm max}=10$ ($\gamma'_{\rm max}=10.05$). For a more realistic value, $\gamma'_{\rm max}\sim10^6$, the orbit is very closely approximated by a triangular form, consisting of a sequence of oppositely directed light lines.}
\label{fig:trajectory}
\end{figure*}

\subsection{Examples}

Examples of the exact solution are shown in Fig.~\ref{fig:trajectory}. A feature of the solutions is that the trajectory is well approximated by straight lines, with one of $z'\mp ct'$ constant, corresponding to $Z'(\chi)\approx\pm\chi$ in the first half phase, with $Z'$ oscillating between approximately $\pm\pi$. The linear approximation is accurate except for a small range of phase, $\sim1/\gamma'_{\rm max}$, where the particles become nonrelativistic and reverse their direction of propagation. As is evident from Fig.~\ref{fig:trajectory}, the trajectory of a background particle ($\beta'_0=0$) involves an oscillation about a fixed point in the primed frame. A test particle with $|u'_0|<|u'_{\rm max}|$, reverses its direction of motion periodically in the LAEW, but for  $|u'_0|>|u'_{\rm max}|$, the direction of motion of the test particle does not change, as its 4-velocity oscillates in the LAEW. For $|u'_0|\gg|u'_{\rm max}|$ the effect of the LAEW on the motion of the test charge may be treated as a perturbation to the motion of a background particle. 

\begin{figure*}
\centerline{
\includegraphics[width=0.4\textwidth]{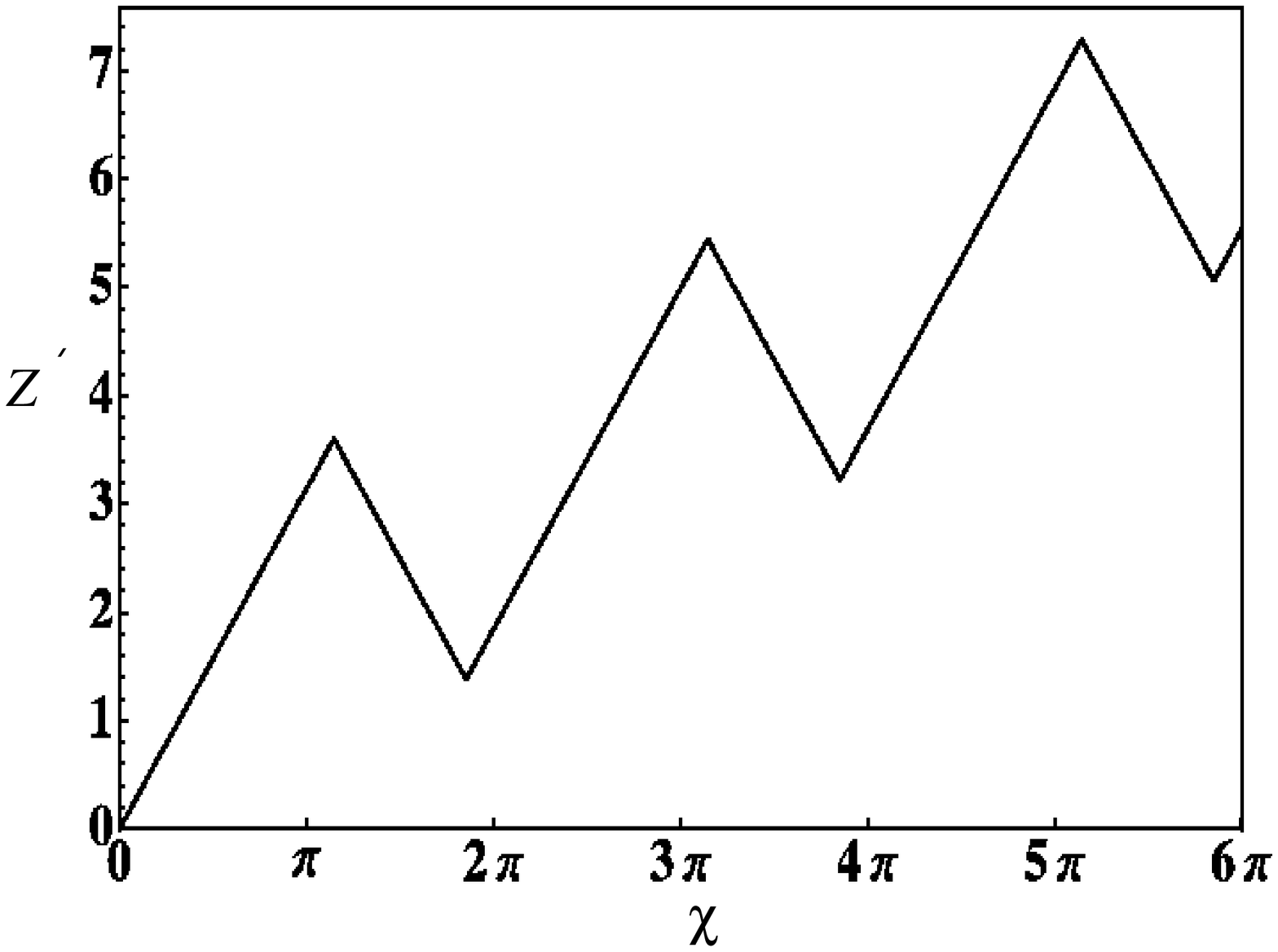}
\qquad
\includegraphics[width=0.4\textwidth]{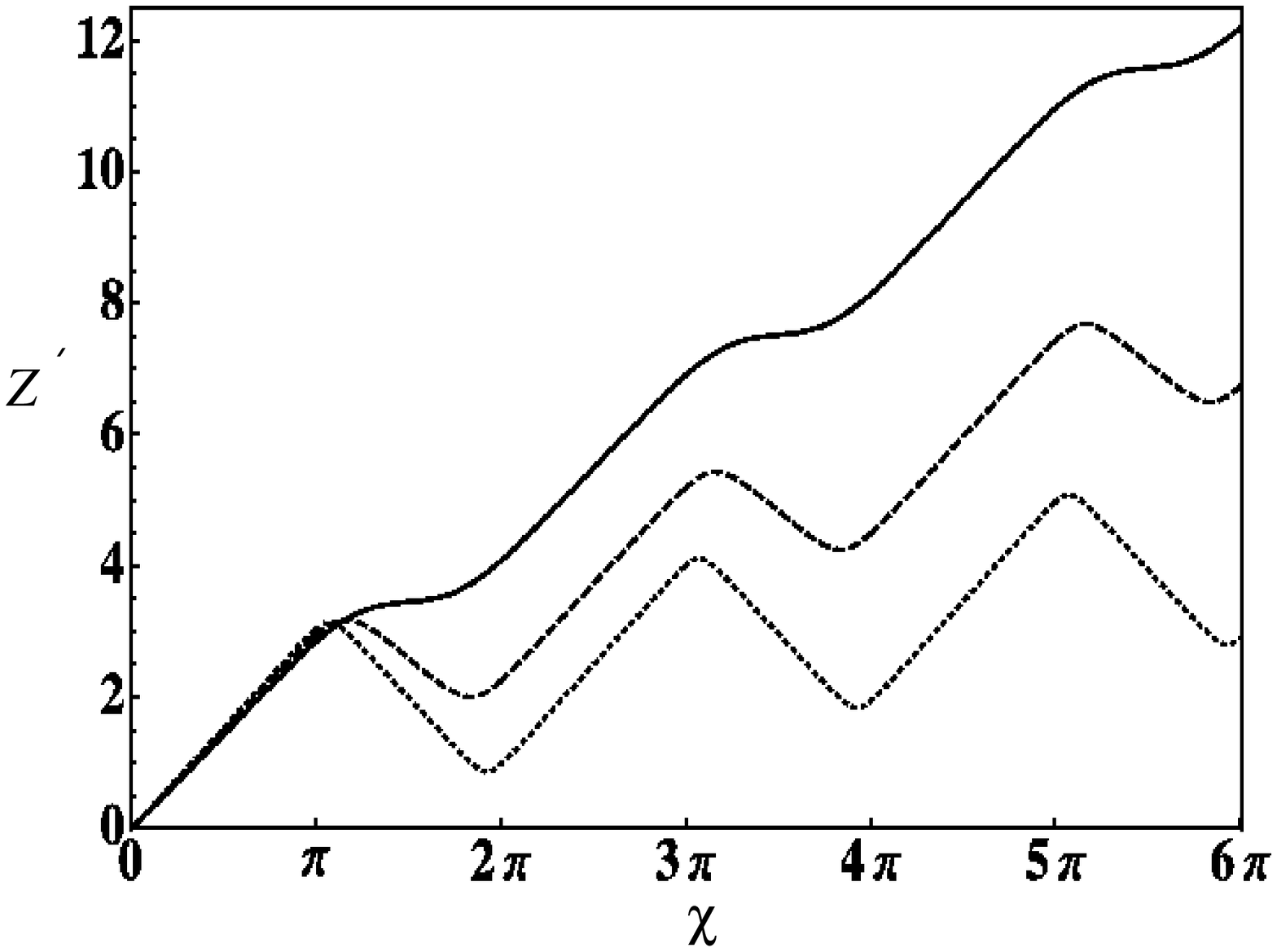}
}
\caption{Trajectory of a test charge, on the left for $ \gamma'_0 \approx 500$, $\gamma'_{\rm max} \approx 1000$, and on the right for $u'_0=1.33$ and for $u'_{\rm max}=5$ (lowest), 2.45 (middle) 1.25 (upper). As in Fig.~\ref{fig:trajectory}, in the highly relativistic case the orbit is well approximated by a sequence of light lines.}
\label{fig:test}
\end{figure*}

For $\gamma'_{\rm max}\gg1$, the orbit of any particle (except for $|\gamma'_{\rm max}-\gamma'_0|\lapprox1$) is well approximated by a sequence of light lines, such that $z'$ varies as $\pm ct'$ in sequential sections, as shown on the left in Fig.~\ref{fig:test}.   For the large value, $\gamma'_{\rm max}\sim10^6$, expected in an oscillating pulsar model, the fraction of the phase, $\Delta\chi\sim1/\gamma'_{\rm max}$, where the particles are nonrelativistic is tiny, and the approximation in terms of a sequence of light lines is very accurate. The distance between the turn-around points for a background particle is very nearly the light distance, $\pi c/\Omega'$, corresponding to the half period in which the particle travels in the same direction. The solution in terms of Weierstrass functions shows how the orbit evolves from one section to another; to illustrate this, an example where the motion is only mildly relativistic is shown on the right in Fig.~\ref{fig:test}.

\section{Particle current associated with LAE}
\label{sect:LAE}

The motivation for the investigation reported is the treatment of LAE due to motion in a LAEW. In Appendix~\ref{sect:covariant} we use the covariant formalism to evaluate the 4-current associated with LAE, and in this section we use the results derived in Appendix~\ref{sect:covariant} to write down the 3-current associated with LAE in the primed frame.

\subsection{LAE in the primed frame}

The emission of LAE can be calculated in terms of the current associated with the motion of the charge in the LAEW. For periodic motion, the current may be expanded in a Fourier series \citep{r95}, giving a sum over a harmonic number $s$. The expansion is written down using a covariant formalism in Appendix~\ref{sect:covariant}. In the primed frame, the current along the direction defined by the LAEW follows from (\ref{Jmu8}):
\be
J_s(\omega',{\bi k}')=-B_s(\omega',{\bi k}')\,2\pi\delta(\omega'-s\Omega'),
\label{Jmu9}
\ee
with $B_s(\omega',{\bi k}')$ following from (\ref{Jmu6}). It follows that in the primed frame, emission at the $s$th harmonic is at the frequency, $\omega'=s\Omega'$. That is, the emission is at harmonics of the frequency of the LAEW in the primed frame. 

The explicit form for the current obtained by inserting (\ref{Jmu6}) into (\ref{Jmu9}) involves an integral over a phase factor, with phase $\phi(\chi)$ say. After replacing the sum over $s$ by an integral and performing the integral over the $\delta$-function in (\ref{Jmu9}), the phase becomes
\be
\phi(\chi)={\omega'\over\Omega'}[\chi-Z'(\chi)\cos\theta'],
\label{phase1}
\ee
in the primed frame. 

On inserting the exact solution (\ref{Zchi}) for the orbit into (\ref{phase1}), it is obvious that some simplifying procedure is needed before one can evaluate the integral over the phase factor. The analogy between LAE and synchrotron emission, discussed in detail in paper~2, suggests that an Airy-integral approximation is appropriate. The Airy integral applies when the phase is approximated by an expansion about a specific phase in which the quadratic term vanishes and only the linear and cubic terms are retained. On making a Taylor expansion of $Z'(\chi)$ about an arbitrary phase, $\chi_0$ say, the quadratic term vanishes when the second derivative of $Z'(\chi)$ is zero at $\chi=\chi_0$. This corresponds to an extremum of $\beta'(\chi)$ and hence of $\gamma'(\chi)$. The only relevant extrema are maxima of $\gamma'(\chi)$, which occur at $\chi_0=\pi/2+n\pi$. Assuming the radiating particle to be highly relativistic near this phase, one may make the approximation $\beta'(\chi)\approx\pm[1-1/2\gamma'^2(\chi)]$. This approximation leads to a highly relativistic counterpart of (\ref{Zchi}). For $u'_0=0$ the integral can be performed exactly, giving
\be
Z'(\chi)=\chi-{1\over8\gamma'^2_{\rm max}}\bigg(
{\chi-\pi/2\over1-4(\chi-\pi/2)^2/\pi^2}
+{\pi\over4}\ln\left|{1+2(\chi-\pi/2)/\pi\over1-2(\chi-\pi/2)/\pi}\right|
\bigg),
\label{TW4}
\ee
for $0<\chi<\pi$. Comparison with (\ref{Zchi}) shows that (\ref{TW4}) is an excellent approximation except near $\chi=0,\pi$ where the particle is briefly nonrelativistic. The Airy-integral approximation is obtained by expanding (\ref{TW4}) around $\chi_0=\pi/2$. The same result is obtained by expanding $\gamma'(\chi)$ around $\pi/2$. The latter procedure is readily generalized to $u'_0\ne0$, for which the maxima in $\gamma'(\chi)$ also occur at $\chi_0\approx\pi/2+n\pi$ and are $\gamma'_\pm\approx\gamma'_{\rm max}\pm u'_0$. The Airy-integral form for the phase (\ref{phase1}) then follows from
\be
Z'(\chi)=\chi-{1\over2\gamma'^2_\pm}\left[
\left(\chi-{\pi\over2}\right)
+{8\over3\pi^2}{\gamma'_{\rm max}\over\gamma'_\pm}\left(\chi-{\pi\over2}\right)^3\right].
\label{TW4a}
\ee
The evaluation of the integral is discussed in detail in paper~2, and here we quote the result and discuss some of its implications.

\subsection{Emission and absorption of LAE}

The formal treatment of LAE involves both conceptual and mathematical complications, discussed in paper~2. To proceed with the discussion here, we quote the results derived in paper~2 for the emissivity (power per unit frequency and per unit solid angle) in the primed frame. In terms of the characteristic frequency $\omega'_{c\pm}=2\Omega'\gamma'^2_\pm$, one has
\be
\eta'_\pm(\omega',\theta')={3q^2\omega'_{c\pm}\gamma'^2_\pm\theta'^2(1+\gamma'^2_\pm\theta'^2)\over16\pi^4\varepsilon_0c}\,
\xi'^2_{c\pm}K^2_{1/3}(\xi'_\pm),
\label{emissivity}
\ee
where $K_{1/3}(\xi)$ is a modified Bessel function, with $\xi'_\pm=\xi'_{c\pm}(1+\gamma'^2_\pm\theta'^2)^{3/2}$, $\xi'_{c\pm}=(\pi/\sqrt{2})(\omega'/\omega'_{c\pm})(\gamma'_\pm/\gamma'_{\rm max})^{1/2}$, and where the $\pm$ signs correspond to emission in the forward (defined for $u'_0>0$) and backward direction. The characteristic properties of LAE implied by (\ref{emissivity}) are discussed in paper~2. 

Let $g'(u'_0)\rmd u'_0$ be the number density of one species of particle, electrons say, with initial (at phase $\chi=0$) 4-velocity in the range $u'_0$ to $u'_0+\rmd u'_0$. The volume emissivity is $\int\rmd u'_0g'(u'_0)\eta'_\pm(\omega',\theta')$. The absorption coefficient can be related to the emissivity (\ref{emissivity}) by an argument using detailed balance \citep{T58,WSW63}. In the one-dimensional case of relevance here this gives
\be
\Gamma'_\pm(\omega',\theta')=-{(2\pi)^3c\over m\omega'^2}\int\rmd u'_0\,\eta'_\pm(\omega',\theta')\,
\cos\theta'{\rmd g(u'_0)\over\rmd u'_0},
\label{growth1}
\ee
with $\cos\theta'\approx\pm1$ such that the signs of $\cos\theta'$ and $u'_0$ are the same. Maser emission corresponds to negative absorption, requiring $\Gamma'_\pm(\omega',\theta')<0$. We comment below on the possibility of maser emission at low frequencies.

As discussed in paper~2, a mathematical inconsistency arises when one attempts to consider the low-frequency limit of (\ref{growth1}) with (\ref{emissivity}). One way of avoiding this difficulty is to consider the average of (\ref{growth1}) over solid angle, denoted by an overline:
\be
{\overline\Gamma}'_\pm(\omega')=-{(2\pi)^3c\over m\omega'^2}\int\rmd u'_0\,{\overline\eta}'_\pm(\omega')\,
{\rmd g(u'_0)\over\rmd u'_0},
\label{growth2}
\ee
where the choice of the sign $\cos\theta'=\pm1$ is implicit. The average of the emissivity is evaluated in paper~2, and at low frequencies for a triangular wave form it reduces to
\be
{\overline\eta}'_\pm(\omega')\approx{2q^2\Omega'\over80\pi^2\varepsilon_0c}
\left({\pi\omega'\over\omega'_{c\pm}}\right)^{4/3}\left({\gamma'_\pm\over\gamma'_{\rm max}}\right)^{2/3}{\rm Ai}^2(0)
\label{eta7t}
\ee
with ${\rm Ai}(0)=1/3^{2/3}\Gamma(2/3)=0.355$. We use (\ref{growth2}) rather than (\ref{growth1}) in the discussion below.

The emissivity (\ref{emissivity}) is in the primed frame, and it is straightforward to transform it to the pulsar frame, in which the LAEW is an outward propagating wave. The Lorentz transformation (\ref{LT2}) implies
\be
\eta_\pm(\omega,\theta)={\omega\over\omega'}\eta'_\pm(\omega',\theta'),
\qquad
\omega=\gamma^*(1\pm\beta^*)\omega',
\qquad
\theta=\gamma^*(1\mp\beta^*)\theta'.
\label{LT3}
\ee
However, it is usually more convenient to discuss the emission and absorption in the primed frame before transforming to the pulsar frame.

\section{Discussion}
\label{sect:discussion}

In this section we discuss four aspects of the foregoing results in further detail: systematic acceleration, the displacement of individual charges in a LAEW, drift motion and its relevance to LAE in a LAEW, and maser LAE in a LAEW.

\subsection{Systematic acceleration}

In solving the equation of motion, we find that a LAEW with an asymmetric wave form, specifically one with $a_0\ne0$ in (\ref{LAE6}), leads to systematic acceleration of all particles. According to (\ref{LAE4}) with (\ref{LAE6}), the 4-velocity increasing by $2\pi a_0$ each wave period. This corresponds to a systematic acceleration by a static electric field ${\bar E}=a_0\Omega'mc/q$. No uniform electric field is included in the analytic model for a LAEW \citep{lm08}, and we assume $a_0=0$ in the foregoing.

It is interesting to speculate on the possible significance of assuming $a_0\ne0$. One could interpret  the ${\bar E}$ associated with $a_0\ne0$ as one way of modeling an unscreened component of the pulsar's inductive field. The systematic acceleration transfers electromagnetic energy into particles. In a conventional polar-cap model, such acceleration occurs in a gap, and screening of the inductive field results from a net charge density in a pair formation front \citep{hm98}. One motivation for an oscillating model is that such screening is unstable to temporal perturbations, resulting in large-amplitude electric oscillations \citep{letal05}. The build up of a LAEW in an oscillating model relies on acceleration by an incompletely screened electric field. As in a stationary model, the resulting pair creation should lead to screening of the electric field, and whereas this occurs locally in a stationary model, it corresponds to a systematic reduction in ${\bar E}$, and hence of $a_0$, in the oscillating model. This effect needs to be included in a detailed theory of the instability leading to the LAEW, but we  do not attempt a quantitative treatment of this here.

\subsection{Displacement of charges in a LAEW}

A particle in a LAEW oscillates about a center that is drifting (except for a background particle in the primed frame). The displacement about this center is by $\pm\pi c/\Omega'$ in the primed frame. A particle is constrained to move along a magnetic field line that is curved, and the neglect of this curvature is valid only if this distance is small compared with the radius of curvature of the field line. For a LAEW with a frequency $\sim10^9\rm\,s^{-1}$, $\pm\pi c/\Omega'$ is less than a meter, and this condition is well satisfied.

A test particle has a drift velocity in the primed frame. This velocity may be estimated from the distance, $\Delta z'$ say, its position advances in each period of the LAEW:
\be
\Delta z'={c\over\Omega'}\int_0^{2\pi}\rmd\chi\,\beta'(\chi),
\label{DeltaZ}
\ee
with $\beta'(\chi)$ given by (\ref{LAE5}). The mean drift velocity is $\Omega'\Delta z'/2\pi$, which depends on $u'_0$, and $u_{\rm max}$, and is zero for $u'_0=0$. The 4-velocity constructed from the 3-velocity $\Omega'\Delta z'/2\pi$ is rather cumbersome. Alternatively, one might consider a drift 4-velocity found by dividing the displacement (\ref{DeltaZ}) by the proper time elapsed in a wave period, $\Delta\tau$ say. One has
\be
\Delta\tau={1\over\Omega'}\int_0^{2\pi}\rmd\chi\,{1\over\gamma'(\chi)},
\label{Deltatau}
\ee
with $\gamma'(\chi)$ given by (\ref{LAE5a}). The ratio $\Delta z'/\Delta\tau$ is some measure of the mean 4-velocity, but describes a different quantity. We note that neither the integral (\ref{DeltaZ}) nor the integral (\ref{Deltatau}) appear in the detailed theory for LAE developed in paper~2.

\subsection{Comparison with \cite{r95}}

The solution found here for the orbit of a particle in a LAEW with a triangular wave form is similar to the solution found by \cite{r95} for a sinusoidal wave form. However, there are important differences in the results for LAE. An important difference is the dependence of the emission formula found by \cite{r95} on a drift velocity, and we comment on this below. First, however, we comment on similarities in the results, to eliminate them as possible reasons for the differences in the treatment of LAE.

As in the present paper, \cite{r95} found that when the particles are highly relativistic over most of the phase, their orbit is well approximated by a sequence of light lines, with the direction of motion reversing at the phase where the particles are briefly nonrelativistic. \cite{r95} also found that the emission of LAE is dominated by the phase where the particles have their maximum Lorentz factor. From one perspective this is a surprising result, which is clearly insensitive to the assumed wave form. Based on the generalized Larmor formula, one expects the emission to be maximum when the acceleration is a maximum, that is, at the phase where the electric field is maximum. In contrast, it is found that the emission of LAE maximizes around the phase where the acceleration is zero. Clearly, the important differences in the treatment of LAE between the present paper and that of \cite{r95} are not due to the assumed wave form, to the detailed form of the orbit or to the phase that determines the properties of the emission. 

\subsection{Absence of Doppler shift in LAE}

In the theory of LAE developed by \cite{r95}, the drift velocity plays an important role, in the sense that the frequency of LAE is at harmonics of the Doppler-shifted frequency of the LAEW. The emission in the theory of \cite{r95} occurs at $(\omega-k_zc\beta_D)-s(\Omega-Kc\beta_D)=0$, or $ku_D-sKu_D$ in invariant form, where $u^\mu_D$ is the drift 4-velocity, $\gamma_D$, $\beta_Dc$ are the associated Lorentz factor and 3-velocity, and $s$ is the harmonic number. In contrast we find no explicit dependence on a drift velocity in our treatment of LAE.  

A drift velocity does appear naturally in the analogous theory for emission due to motion in a large amplitude transverse wave (LATW) \citep{go71,a72}. In a LATW the oscillatory motion is in the transverse plane and there is a drift motion along the direction of wave propagation. The frequency of emission by such a particle is at harmonics of the Doppler-shifted wave frequency, with the Doppler shift corresponding to the drift velocity. However, the analogy between the transverse and longitudinal cases is misleading when considering the emission of radiation. Semi-quantitatively, one may regard the motion in a LATW as consisting of two effectively independent motions: that in the transverse plane causing the radiation, and that along the axis providing the Doppler shift. In a LAEW both the drift and the oscillatory motion are along the axis, and the two motions cannot be separated in this way. 

No Doppler shift associated with the drift motion of a test particle appears explicitly in the phase (\ref{phase1}). As shown in Appendix~\ref{sect:covariant}, explicit inclusion of the drift motion involves extracting a term, $a_1\chi$ from $Z'(\chi)$, such that ${\tilde Z}'(\chi)=Z'(\chi)-a_1\chi$ is periodic. This does introduce a Doppler shift, as assumed by \citep{r95}. In the continuum approximation, when $s$ is replaced by a continuous variable, the sum over $s$ is replaced by an integral that is performed over the resonance condition, $s$ is replaced by the Doppler shifted frequency divided by the frequency of the LAEW. The phase is given by (\ref{phase1}) irrespective of whether the Doppler shift is included or not. If the Doppler shift is ignored, one has $s=\omega'/\Omega'$ and (\ref{phase1}) follows by the argument given; if the Doppler shift is included, the term $a_1\chi$ is subtracted from $Z'(\chi)$, added to $s$, and restored to ${\tilde Z}'(\chi)$ to give $Z'(\chi)$ after integration of $s$. When the integral of the phase factor over $\chi$ is evaluated in terms of an Airy integral, as in paper~2, no separation of the term $a_1\chi$ associated with the drift is necessary or appropriate. It follows that no Doppler shift appears explicitly in LAE.

A physical explanation for the absence of an explicit Doppler shift in LAE (in the primed frame) is as follows. The emission of LAE is concentrated around the phase where $\gamma'(\chi)$ reaches a maximum, with $\rmd\gamma'(\chi)/\rmd\chi=0$. This is the phase where the acceleration reverses sign and is instantaneously zero; all particles have their maximum Lorentz factor at the same phase. The theory of LAE depends only on the maximum value of $\gamma'(\chi)$ and the way that it varies with $\chi$ around this maximum. The emission does depend on the initial 4-velocity, $u'_0$, which also determines the drift motion, but the two dependences on $u'_0$ are quite different. In the emission of LAE, $u'_0$ affects the shape of the variation of $\gamma'(\chi)$ about its maximum. In contrast, the drift velocity is determined by the integral (\ref{DeltaZ}) for the displacement in a period of the LAEW, and this is dominated by the asymmetry between the forward and backward paths, as shown in Fig.~\ref{fig:test}; in this case, $u'_0$ affects the turn-around points, where $\gamma'(\chi)$ passes through its minimum of unity. The dependence of LAE on $u'_0$ cannot be described in terms of the drift velocity.

\subsection{Maser LAE at low frequencies}

Maser LAE is known to be possible for small amplitude oscillations, when the frequency of the maser is determined by a characteristic frequency in the radio range \citep{m78}. Here the characteristic frequency, $\omega'_{c\pm}\sim2\Omega'\gamma'^2_{\rm max}$, is at a much higher frequency, which is estimated in paper~2 to be in the hard X-ray range for the largest-amplitude LAEWs. For maser LAE in a LAEW to be relevant as a pulsar emission radio emission mechanism, the maser needs to operate at low frequencies, $\omega\ll\omega_{c\pm}$. It follows from (\ref{growth1}) that negative absorption requires two conditions be satisfied. First, the distribution function must satisfy $\rmd g(u'_0)/\rmd u'_0>0$, over at least some range of $u'_0$. This condition is plausibly satisfied for pairs generated in a cascade. Numerical models for pair creation \citep{zh00,ha01,ae02} suggest that the distribution of particles increases with energy with a peak between $\gamma'_0$ of a few to $\sim10^2$. **These calculations are for a stationary model, and we assume that the results are similar for the oscillating model, such that the intrinsic spread in initial Lorentz factors is $\sim10^2$. This spread is small compared with $\gamma_{\rm max}\sim10^6$--$10^7$. Due to this spread, it is plausible that the initial distribution function is an increasing function of $u'_0$ with a peak in the range 1--$10^2$, implying $\rmd g(u'_0)/\rmd u'_0>0$ in (\ref{growth1}).**  Second, by partially integrating (\ref{growth1}), one finds that $\rmd\eta'_\pm(\omega',\theta')/\rmd u'_0<0$ is also required for negative absorption to be possible. As already remarked, applying this condition to the emissivity itself encounters inconsistencies, and we avoid these by considering the angle-averaged form (\ref{growth2}). Using (\ref{eta7t}), one finds ${\overline\eta}'_\pm(\omega')\propto1/\gamma'^2_\pm$, and this requirement is well satisfied.

The growth rate for LAE follows by partially integrating (\ref{growth2}) and using (\ref{eta7t}):
\be
{\overline\Gamma}'_\pm(\omega')=-{2(2\pi)^3c\over m\omega'^2}\int\rmd u'_0\,g'(u'_0)\,
{{\overline\eta}'_\pm(\omega')\over\gamma'_\pm},
\label{growth3}
\ee
where $|\rmd\gamma'_\pm/\rmd u'_0|=1$ is used. Assuming $u'_0$ is negligible in comparison with $\gamma'_{\rm max}$, one finds
\be
{\overline\Gamma}'_\pm(\omega')=-{\pi^{7/3}\over2^{1/3}}\left({\Omega'\over\omega'}
\right)^{5/3}\left({1\over\gamma'_{\rm max}}\right)^{8/3}.
\label{growth4}
\ee
The theory is valid only for $\omega'\gg\Omega'$ and $\gamma'_{\rm max}\gg1$, so that both factors on the right hand side are very small. It follows that although the condition for growth  at low frequency are well satisfied, the growth rate is very small. To explain pulsar radio emission in terms of maser LAE in a LAEW one would need LAEWs with much smaller amplitudes, with $\gamma'_{\rm max}\lapprox10$. 

\section{Conclusions}
\label{sect:conclusions}

In this paper, we solve the equation of motion for a charged particle in a large amplitude electrostatic wave (LAEW), both in an arbitrary frame (Appendix~\ref{sect:covariant}) and in the (primed) frame in which the oscillations are purely temporal (Sec.~\ref{sect:pframe}). The case of interest for pulsars is when the electrons and positrons are accelerated to highly relativistic energies in a small fraction of a period of the wave, and in this case the wave form of the LAEW is well approximated by the triangular form Fig.~\ref{fig:efield}. The orbit of a particle is described in terms of its displacement, $z$, along the axis defined by the LAEW, as a function of phase $\chi$. We choose the zero of the phase such that the maxima and minima of the wave form for the electric amplitude at $\chi=n\pi$, with $n$ an integer.

We distinguish between background particles and test particles. Background particles are assumed to come to rest at the maxima and minima in the wave form, and reach a common maximum Lorentz factor, $\gamma'_{\rm max}$ at the intermediate phases, $\chi=\pi/2+n\pi$. A test particle is any particle with a nonzero 4-velocity ($u'_0=\gamma'_0\beta'_0$, $\gamma'_0=(1-\beta'^2_0)^{-1/2}$) at the initial phase, $\chi=0$. The Lorentz factor of a test charge oscillates between maxima $\gamma'_\pm\approx|\gamma'_{\rm max}\pm u'_0|$ at $\chi=\pi/2+n\pi$; a test charge comes to rest for $\gamma'_0<\gamma'_{\rm max}$ twice per period, at phases that depend on $u'_0$.  We derive an explicit expression for the orbit of a particle in a LAEW with a triangular wave form in terms of elliptic integrals (Weierstrass functions). In the highly relativistic case, the exact solution for the orbit is well approximated by a set of light lines, with the velocity changing rapidly between $\pm c$ at phases $\chi=\pi/2+n\pi$.  This leads to a triangular form for the orbit of a background particle, which oscillates about a fixed point. A test charge oscillates about a moving center, with the drift speed of the center of oscillation depending on $\beta'_0$, but there is no simple analytic expression for the drift speed. 

Our motivation for this investigation was to explore the properties of linear acceleration emission (LAE) in a LAEW, particularly the possibility of maser LAE in a LAEW as a radio emission mechanism, and high-energy LAE as an intermediate step in a pair cascade. The treatment of LAE involves some conceptual and mathematical difficulties and is discussed separately in paper~2 \citep{ml09}. Here we write down the emissivity derived in paper~2, and use it to consider the absorption coefficient for LAE, with negative absorption implying maser emission. A mathematical difficulty is avoided by averaging the absorption coefficient over the small cone of emission. Maser emission is found to be possible for LAE in LAEW, as in earlier theories: \cite{m78} found maser LAE in a small amplitude electrostatic oscillation with $\gamma'_0\gg\gamma'_{\rm max}$ in the present notation, and \cite{r95} found maser LAE in a theory for LAE in a sinusoidal LAEW. We note that the latter theory \citep{r95} depends explicitly on a drift velocity, and in this sense is incompatible with the theory developed here and in paper~2. We discuss this difference and explain why there should be no explicit dependence on a drift velocity in LAE. 

Our estimate of the growth rate for maser LAE at low frequencies implies that it is too small for effective growth in a LAEW with an amplitude needed to trigger pair creation in an oscillating model for pulsars. For LAE in a LAEW to be viable as a radio emission mechanism, one requires a smaller amplitude LAEW, with $\gamma'_{\rm max}\lapprox10$. It seems plausible that LAEWs with a wide range of amplitudes are present in different regions of the magnetosphere, and one needs to assume this to be the case for maser LAE to be effective. We conclude that maser LAE in a LAEW is a possible radio emission mechanism, but that the mere presence of a LAEW does not ensure effective maser LAE. We discuss the possible relevance of LAE to high-energy processes in pulsars in paper~2. 

\acknowledgements

We thank Matthew Verdon for helpful comments on the manuscript.

\begin{appendix}

\section{Motion in a LAEW: covariant treatment}
\label{sect:covariant}

In this section we solve the equation of motion to find the orbit of a charge in a LAEW using a covariant formalism.

\subsection{Covariant formalism}

In the covariant formalism used here, greek indices run over 0--3, with signature $+,-,-,-$, and units with $c=1$ are adopted. An electromagnetic field is described by the Maxwell tensor, $F^{\mu\nu}$, and for a LAEW with a phase $\chi$, which is an invariant, this is of the form $F^{\mu\nu}=E(\chi)f^{\mu\nu}$, where $E(\chi)$ is the invariant electric field defined by $2E^2(\chi)=-F^{\mu\nu}F_{\mu\nu}$. For a LAEW along the 3-axis, the tensor $f^{\mu\nu}$ has components $f^{03}=-1$, $f^{30}=1$, $f^{\mu\nu}=0$ otherwise. The motion of the particle is described by its 4-momentum $p^\mu=mu^\mu$, where $m$ is its rest mass and $u^\mu$ is its 4-velocity. Newton's equation of motion, in covariant form for the Lorentz force, is
\be
m{\rmd u^\mu(\tau)\over\rmd\tau}=qF^{\mu\nu}(\chi)u_\nu(\tau),
\label{A1}
\ee
where $\tau$ is the proper time of the particle. The orbit may be written in the covariant form
\be
x^\mu=x_0^\mu+X^\mu(\tau),
\qquad
u^\mu(\tau)={\rmd X^\mu(\tau)\over\rmd\tau},
\label{orbit1}
\ee
where $x^\mu=[t,{\bi x}]$ is the 4-position and $x_0^\mu$ contains the initial conditions. Our objective is to integrate (\ref{A1}) once to find $u^\mu(\tau)$, and to integrate a second time to find $X^\mu(\tau)$. The LAEW is described by the wave form (\ref{Tchi}), with a wave 4-vector, $K^\mu=[\Omega,{\bi K}]$. where $\Omega$ is its frequency,  and its wave vector, ${\bi K}$, is assumed along the 3-axis. The  invariant $K^2=K_\mu K^\mu=\Omega^2-|{\bi K}|^2$ satisfied $K^2>0$ for a superluminal LAEW. 

The one-dimensional motion is in the two-dimensional 0-3 subspace. The equation of motion then reduces to two invariant components, identifying by choosing two orthogonal 4-vectors that span this subspace. We choose the 4-vectors $K^\mu$ and $K_D^\mu=f^{\mu\nu}K_{\nu}$, which are orthogonal, $K_D^\mu K_\mu=0$. The invariants $K^2$,  $Ku(\tau)=K^\mu u_\mu(\tau)$, $K_Du(\tau)=K_D^\mu u_\mu(\tau)$ are related by
\be
K^2=-K_D^2=[Ku(\tau)]^2-[K_Du(\tau)]^2.
\label{K2}
\ee 
The projection on the 4-velocity and the orbit onto these two 4-vectors gives
\be
u^\mu(\tau)={Ku(\tau)\,K^\mu-K_Du(\tau)\,K_D^\mu\over K^2},
\qquad
X^\mu(\tau)={KX(\tau)\,K^\mu-K_DX(\tau)\,K_D^\mu,
\over K^2}.
\label{A2}
\ee
respectively. An important step involves writing the phase of the LAEW in terms of these invariants. The phase is the invariant $Kx$, evaluated along the (\ref{orbit1}), so that the electric field in the equation of motion (\ref{A1}) is that at the position of the particle. Ignoring a constant in the phase, it is identified as
\be
\chi(\tau)=KX(\tau),
\qquad
{\rmd\chi(\tau)\over\rmd\tau}=Ku(\tau).
\label{phase}
\ee
Using (\ref{phase}), the independent variable in (\ref{A1}) is changed from $\tau$ to $\chi$, and all quantities are regarded as functions of $\chi$.

Projecting (\ref{A1}) onto the 4-vectors $K^\mu$, $K_D^\mu$ gives
\be
{\rmd Ku(\tau)\over\rmd\chi}=\epsilon\omega_ET(\chi){ K_Du(\tau)\over  Ku(\tau)},
\qquad
{\rmd K_Du(\tau)\over\rmd\chi}=\epsilon\omega_ET(\chi),
\label{A3}
\ee
where we use (\ref{Tchi}), and where the dependence of $\chi$ on $\tau$ is implicit.
 
Integrating the second of (\ref{A3}) gives
\be
K_Du(\tau)=K_Du_0+\omega_EF(\chi),
\label{A4a}
\ee
where $K_Du_0$ is a constant of integration, and with
\be
F(\chi)=\epsilon\int_0^\chi\rmd\chi'\,T(\chi').
\label{LAE10a}
\ee 
The solution of the first of (\ref{A3}) is
\be
Ku(\tau)=\{K^2+[K_Du(\tau)]^2\}^{1/2},
\label{A4}
\ee
with only the positive square root allowed for $K^2>0$.

The function $F(\chi)$ may be separated into a periodic part and a systematically increasing part,
\be
F(\chi)={\tilde F}(\chi)+a_0\chi,
\qquad
a_0={\epsilon\over2\pi}\int_0^{2\pi}\rmd\chi\,T(\chi),
\label{LAE6}
\ee
with ${\tilde F}(\chi+2\pi)={\tilde F}(\chi)$. The parameter $a_0$ is zero for any wave form that is symmetric, in the sense $T(\chi+\pi)=-T(\chi)$, and we restrict our discussion to such wave forms. The significance of $a_0\ne0$ is discussed in \S\ref{sect:discussion}. 

The orbit is found by integrating ${\rmd X^\mu(\tau)/\rmd\tau}=u^\mu(\tau)$, which becomes
\be
{\rmd X^\mu(\tau)\over\rmd\chi}={K^\mu\over K^2}-{K_Du(\tau)\over K^2 Ku(\tau)}K_D^\mu
={u^\mu_0+{\tilde u}^\mu(\chi)\over Ku(\chi)}.
\label{A5}
\ee
Integrating the first form in (\ref{A5}) gives
\be
X^\mu(\tau)=x_0^\mu+{\chi K^\mu-Z(\chi)K_D^\mu\over K^2},
\label{A6}
\ee
where an invariant, dimensionless form of the displacement is given by
\be
Z(\chi)=\int_0^\chi\rmd\chi'{K_Du_0+\omega_EF(\chi')\over
\{K^2+[K_Du_0+\omega_EF(\chi')]^2\}^{1/2}},
\label{A8}
\ee
where (\ref{A4a}) and  (\ref{A4}) are used. The proper time of the particle depends on phase through
\be
\tau(\chi)=\int_0^\chi\rmd\chi'{1\over
\{K^2+[K_Du_0+\omega_EF(\chi')]^2\}^{1/2}}.
\label{A9}
\ee

\subsection{Covariant form for the 4-current}

The source term for emission of radiation by any particle is identified as the Fourier transform of the current density associated with the particle \citep{mmcp}. In a covariant formulation, this becomes the 4-current density 
\be
J^\mu(k)=q\int\rmd\tau\,u^\mu(\tau)\,\rme^{\rmi kX(\tau)}.
\label{Jmu1}
\ee
For motion in a LAEW, the only nonzero components of $J^\mu(k)$ are in the 0-3~plane. The formalism introduced in (\ref{A2}) allows one to write
 \be
J^\mu(k)={KJ(k)\,K^\mu-K_DJ(k)\,K_D^\mu\over K^2}.
\label{Jmu2}
\ee
Using (\ref{A6}), the phase factor in (\ref{Jmu1}) becomes
 \be
kX(\tau)=kx_0+\chi\,{kK\over K^2}+Z(\chi)\,{kK_D\over K^2}.
\label{Jmu3}
\ee
Using (\ref{A2}), (\ref{Jmu1}) gives
\be
\left[
\!\!
\begin{array}{c}
KJ(k)\\
K_DJ(k)
\end{array}
\!\!
\right]=q\int\rmd\chi\,\left[
\!\!
\begin{array}{c}
1\\
\beta(\chi)
\end{array}
\!\!
\right]\,\rme^{\rmi kX(\chi)},
\label{Jmu4}
\ee
with $\beta(\chi)=K_Du(\chi)/Ku(\chi)$, and with $ kX(\chi)$ now regarded as a function of $\chi$.

The integral in (\ref{Jmu4}) may be evaluated after expanding in a Fourier series \citep{r95}. For a background particle, $Z(\chi)$ is a periodic function and the expansion in Fourier series may be written
\be
\left[
\!\!
\begin{array}{c}
1\\
\beta(\chi)
\end{array}
\!\!
\right]\,\rme^{\rmi Z(\chi)\,kK_D/K^2}
=\sum_{s=-\infty}^\infty
\left[
\!\!
\begin{array}{c}
U_s(k)\\
B_s(k)
\end{array}
\!\!
\right]\,\rme^{-\rmi s\chi},
\label{Jmu5}
\ee
with the Fourier coefficients given by
\be
\left[
\!\!
\begin{array}{c}
U_s(k)\\
B_s(k)
\end{array}
\!\!
\right]
=
{1\over2\pi}\int_0^{2\pi}\rmd\chi\,
\left[
\!\!
\begin{array}{c}
1\\
\beta(\chi)
\end{array}
\!\!
\right]\,\rme^{\rmi[s\chi+Z(\chi)\,kK_D/K^2]}.
\label{Jmu6}
\ee
Then (\ref{Jmu4}) becomes
\be
\left[
\!\!
\begin{array}{c}
KJ(k)\\
K_DJ(k)
\end{array}
\!\!
\right]=\sum_{s=-\infty}^\infty \left[
\!\!
\begin{array}{c}
KJ_s(k)\\
K_DJ_s(k)
\end{array}
\!\!
\right],
\label{Jmu7}
\ee
with the Fourier coefficient determined by
\be
\left[
\!\!
\begin{array}{c}
KJ_s(k)\\
K_DJ_s(k)
\end{array}
\!\!
\right]=q\left[
\!\!
\begin{array}{c}
U_s(k)\\
B_s(k)
\end{array}
\!\!
\right]2\pi\delta(s-kK/K^2).
\label{Jmu8}
\ee

For a test particle, one may write $Z(\chi)=a_1\chi+{\tilde Z}(\chi)$, with a linear term, $a_1\chi$ associated with the drift motion, and with ${\tilde Z}(\chi)$ a periodic function. The Fourier series may be defined in terms of coefficients ${\tilde U}_s(k)$, ${\tilde B}_s(k)$ defined by (\ref{Jmu6}) with $Z(\chi)\to{\tilde Z}(\chi)$. The result (\ref{Jmu8}) is modified by $U_s(k),B_s(k)\to{\tilde U}_s(k),{\tilde B}_s(k)$ and $s\to s-a_1$ in the $\delta$-function. After replacing the sum over $s$ by an integral, performed over the $\delta$ function, the result (\ref{phase1}) is found to apply to both background and test particles. The separation of $Z(\chi)$ into a drift term and an oscillatory term is irrelevant when treating LAE.

\end{appendix}

\end{document}